\documentstyle[a4,11pt,amsfonts,epsfig,verbatim]{article}
\newcommand {\be} {\begin{equation}}
\newcommand {\ee} {\end{equation}}                                         
\newcommand {\bea} {\begin{eqnarray}}
\newcommand {\eea} {\end{eqnarray}}

\newcommand{\Tr}{\mathop{\mathrm{Tr}}}
\newcommand{\Dtau}{\mathop{\mathrm{\Delta \tau}}}
\newcommand{\Nt}{\mathop{\mathrm{N_t}}}
\newcommand{\Lx}{\mathop{\mathrm{L}}}
\newcommand{\D}{\mathop{\mathrm{D}}}

\newcommand{\id}{{\rm 1\kern-.12em
\rule{0.3pt}{1.5ex}\raisebox{0.0ex}{\rule{0.1em}{0.3pt}}}\,}

\begin{document}

\title{
Monte Carlo study of the magnetic properties 
of the 3{\rm {D}} Hubbard Model}
 
\author{Isabel Campos and James W. Davenport}
\maketitle

\begin{center}
{\small 
  {Brookhaven National Laboratory} \\
{\sl Center for Data Intensive Computing}, \\
 {\rm Upton, 11973 New York (USA)} \\
 e-mail: \tt isabel@sol.unizar.es, daven@bnl.gov} \\
\end{center}

\bigskip
 
\begin{abstract}
We investigate numerically the magnetic properties of 
the $3{\rm {D}}$ Isotropic and Anisotropic Hubbard model at half-filling. 
The behavior of the transition temperature as a function of
the anisotropic hopping parameter is qualitatively described.
In the Isotropic model we measure the scaling properties 
of the susceptibility finding agreement  
with the magnetic critical exponents of the $3{\rm {D}}$ Heisenberg
model. We also describe several particularities concerning
the implementation of our simulation in a cluster of personal computers.

\end{abstract}

\newpage

%
%
%
%

\section{Introduction}

Many electron systems have been for many years the playground for testing
models of superconductivity. In this framework the Hubbard model
\cite{Hubbard1963} is expected to reproduce the essential 
features that purely electronic degrees of freedom are accountable for.
In the Hubbard Hamiltonian thermal agitation is modeled via an 
inter-orbital hopping term, while electrostatic interaction is taken 
into account via an {\sl {effective}} Coulomb coupling proportional 
to the charge density.

The simplicity of this formulation is only apparent.
The Hubbard model has (partial) exact solutions
only in one dimension. In two or more dimensions only approximate
techniques can be applied. Ground state properties and
approximate phase diagrams have been analytically derived in the
limits $\rm{U\rightarrow 0}$ and $\rm{U\rightarrow \infty}$
by using Random Phase Approximation and Strong Coupling expansions 
respectively. It is clear that
the range of validity of analytical approaches 
is always an issue, moreover
we are interested in the physics of the system at
the (more physical) intermediate values of the coupling. 
In those regions where strong and weak coupling expansions break down
Quantum Monte Carlo (QMC) techniques are useful, not only 
for testing the validity of other analytical methods such as Mean Field, 
but as a powerful tool to obtain first-principle results.

The numerical investigation of the Hubbard model is however very costly.
Away from half-filled bands the path integral measure is no longer
positive-definite and Monte Carlo averages of physical observables 
suffer from large fluctuations. Extracting meaningful results
requires extremely large statistics.  
Even in half-filled bands the complexity of the numerical 
simulation is high, as we shall discuss throughout the paper. 
One has to keep in mind that
we are dealing with fermionic particles in a region of parameter
space in which the Coulomb interaction is big enough for the
fermion propagator to be close to singular. Inverting the propagator
is a computationally very expensive operation.
The most popular algorithm \cite{White1989},
the determinantal method, has a computational complexity which
grows with the cube of the system size. Therefore while
the system has been studied extensively in two dimensions, in
three dimensions the information we have is more restricted.

We decided to investigate in this work the magnetic properties of the
three dimensional Hubbard model at half-filling around 
the N\'eel phase transition.
We concentrate on two issues: the transition temperature, 
and the universality class of the transition. Pioneering 
works \cite{Hirsch1987,Scalapino1989}
on the phase diagram were carried out in the 1980s on small lattices.
The model presents a phase transition line in the plane $\beta \, - \, \rm{U}$
separating a region in which the system is in a disordered paramagnetic phase
from another region in which the electron spins are aligned in a staggered
way. The actual value of the transition temperature  
as a function of \rm{U} remains an open problem. Recent numerical simulations
\cite{Muramatsu2000} have shown that 
the phase transition takes place at temperatures much lower than 
expected from earlier works. Another question to be discerned numerically
is the universality class of the phase transition line. 
Close to the Quantum Critical Point at $\rm{T=0}$ the
transition should be mean field like ($\rm{4D}$ Heisenberg exponents). 
Everywhere else it is expected to be in the universality class of the 
three dimensional quantum Heisenberg antiferromagnet \cite{Sandvik1998}.

In recent years the development of clusters of commercial processors
has boosted the computing capabilities at research institutes.
This low weight approach to high speed computing is becoming
a real alternative to more conventional approaches based on parallel
supercomputers traditionally developed by industry. 
While price is an obvious advantage in favor of {\em {hand made}}
clusters of PCs, effectiveness is both a model and algorithm dependent issue.

Obvious cases for massive cluster simulation are problems that, while
not requiring a huge amount of memory, do require the simulation of many 
copies of the same program. 
The speeding up is achieved by setting up different starting
conditions for the different copies of the program. 
On single processors, high performances can be achieved by using the
capabilities of recently developed processors, such as vectorization
\cite{Martin2001}. In this situation the so-called {\em {farm method}}
on clusters of PCs is a cost-effective, relatively easy to use, source of 
computer power. The previous statement is only partially true in our case.
It is not obvious that the {\em {farm method}} is the best solution for the
Hubbard model. Warm up times in large lattices
might be very large and to have thermalization effects under control
it is desirable to have a long single Monte Carlo history.

These and related questions will be addressed here. Simulations
have been carried out on the PC cluster at the Center for Data 
Intensive Computing in Brookhaven National Laboratory.
It consists of about 150 Pentium III  processors. Clock speeds range from
500 MHz to 1GHz,  and the available DRAM is 1 Gbyte per processor.
Communication between the processors is achieved 
via a commercial Fast Ethernet switch which provides a 
one-way bandwidth per channel of up to 100 Mbits/sg.

\vspace{0.5cm}
The paper is organized as follows: we first briefly review the
model and algorithm in section 2 in order to fix our notations; 
the numerical simulation, observables and dynamics of the Monte Carlo 
process is discussed in section 3. Our results for the Isotropic 
and Anisotropic Hubbard model are described in sections 4 and 5 respectively.
Section 6 contains conclusions and comments about future perspectives.
We left for the appendix some particularities related to our
implementation of the simulation in a cluster of PCs.

%
%
%
%

\section{Model and Numerical Algorithm}

In the following we summarize the standard
procedure to perform Monte Carlo simulations on the Hubbard Model
\cite{Blancken1981,White1989,Gubernatis1991} and fix our notation.
Consider the Hubbard Hamiltonian at half-filling 
\be
{\rm {\hat{H}}} = -\rm{t} \, \sum_{\langle \rm{i\,j} \rangle, \alpha} \,
({\rm {c^{\dagger}_{\rm{i}\alpha}}} {\rm {c_{\rm{j}\alpha}}} +
  {\rm {c^{\dagger}_{\rm{j}\alpha}}} {\rm {c_{\rm{i}\alpha}}})
+ {\rm {U}} \, \sum_{\rm{i}} ({\rm {n_{\rm{i}+}}} - 1/2) \, 
({\rm {n_{\rm{i}-}}} - 1/2) \equiv {\rm {\hat{K}}} + {\rm {\hat{V}}}  \ . 
\ee
Here the index ${\rm {i}} = 0, \dots , \rm{V} \equiv \Lx^{\D}$ 
labels the sites of a lattice 
of side $\Lx$ in $\D$ spatial dimensions on which
periodic boundary conditions have been imposed;
$\rm{c}_{\rm{i}\alpha}^{\dagger}$ and  
$\rm{c}_{\rm{i}\alpha}$ are respectively the creation and
annihilation operators for electrons with a z-component of spin $\alpha$ at
the site ${\rm {i}}$; $\rm{n}_{{\rm {i\alpha}}} 
= \rm{c}_{\rm{i}\alpha}^{\dagger} \rm{c}_{\rm{i}\alpha} $ denotes the 
usual number operator.
The sum $\langle \rm{i \, j} \rangle$ is over all pairs 
of nearest neighbors
on the lattice. The first term models the thermal agitation, 
whose strength is characterized by the hopping parameter ${\rm {t}}$; 
the second term corresponds to an
electrostatic Coulomb repulsion of intensity ${\rm {U}}$.

The path-integral representation is obtained by introducing an imaginary
time coordinate $\tau$, and considering the action of all possible 
configurations of the fields between $\tau=0$ and $\tau = \beta$. 
The partition function of the equivalent statistical model reads 
\begin{equation}
{\mathcal {Z}} \equiv \Tr \, 
(\rm{e}^{-\hat{\rm{S}}}) = \Tr \, (\rm{e}^{-\beta \hat{\rm{H}}}) \ .
\label{Z}
\end{equation}

In order to perform a numerical simulation the theory is defined on a lattice
in space and time dimensions. The partition function of the discretized
theory can be written as
\be
{\mathcal {Z}} =
\Tr \, ( \rm{e}^{-\Delta\tau \rm{\hat{K}} - \Delta\tau \rm{\hat{V}}} )^{\Nt} \ .
\label{ZZ}
\ee
where we defined $\beta \equiv \Dtau \, \Nt$, with
$\Dtau$ lattice spacing in the temporal direction and $\Nt$
number of time slices. 
The role of $\beta$ is analogous to the inverse of the temperature $\rm{T}$
of a classical statistical model in $\rm{D+1}$ dimensions.

In order to have a well defined relative probability for each 
configuration in phase space, fermions must be integrated out analytically in
the partition function (\ref{ZZ}). In a first step the kinetic and potential
terms are separated in the partition function by the splitting 
\begin{equation}
{\mathcal Z} = \Tr \, (\rm{e}^{-\Delta\tau \rm{\hat{K}} }
\, \rm{e}^{-\Delta\tau  \rm{\hat{V}}  } )^{\Nt} + 
{\mathcal O}(\Delta\tau^2 [\hat{\rm{K}},\hat{\rm{V}}] )  \ .
\label{trotter}
\end{equation}
We will neglect the contributions coming from the second term on the
r.h.s of eq. (\ref{trotter}). 
The leading order in the error introduced by the so-called Trotter aproximation
is proportional to the square of $\Dtau$. 
This systematic error has to be kept under control
in actual numerical simulations by choosing $\Dtau$ small enough (i.e.
smaller than the statistical error).

The kinetic part of the Hamiltonian is a quadratic form in fermion fields,
calculating the trace is therefore trivial. In order to  
write the interaction term as a quadratic form as well one introduces
a set of auxiliary boson fields \cite{Hirsch1983}
\begin{eqnarray}
\rm{e}^{-\Delta\tau \, U (\rm{n_{i+}} - 1/2)(\rm{n_{i-}} - 1/2)} =
\frac{\rm{e}^{-\Delta\tau \, \rm{U/4}}} {2} 
\sum\limits_{\sigma_{\rm{i}}(\rm{l}) = \pm 1}
\rm{e}^{-\Delta\tau \, \sigma_{\rm{i}}(\rm{l}) \, \lambda 
(\rm{n_{i+}} - \rm{n_{i-}})} \ ,
\label{ising}
\end{eqnarray}
where $\{ \sigma \}(\rm{l})$ denotes an Ising field defined 
in the spatial lattice at the time slice
${\rm {l=1, \dots , \Nt}}$. 
The constant $\lambda$ is related to the parameters of the Hamiltonian by
the equation ${\rm{cosh}} (\Dtau \lambda) = \exp(\Dtau  \rm{U/2})$
for positive $\lambda$.

After all these manipulations it is possible to perform the trace 
in (\ref{Z}) yielding
\begin{equation}
{\mathcal Z} = \sum\limits_{ \{ \sigma(\rm{l}) \}} \,
\det \hat{\rm{M}}^{+} \, \det \hat{\rm{M}} ^{-} \ ,
\label{partition}
\end{equation}
with
\begin{equation}
\hat{\rm{M}}^{\alpha} = \id +
\hat{\rm{B}}_{\rm{N_t}}^{\alpha} \hat{\rm{B}}_{\rm{N_t -1}}^{\alpha} \cdots 
\hat{\rm{B}}_1^{\alpha}
\equiv \id + \hat{\rm{A}}^{\alpha}(\Nt)  \ .
\end{equation}
where we have defined the matrices 
\begin{eqnarray}
\hat{\rm{B}}_{\rm{l}}^{\alpha} & = &
\rm{e}^{\mp \lambda \, \Delta\tau \, \alpha \, \delta_{\rm{ij}} \,
\sigma_{\rm{i}}(\rm{l})} \, \rm{e}^{- \Delta\tau \, \hat{\rm{K}}}  \\[2ex]
\hat{\rm{A}}^{\alpha}(\rm{l}) & = & 
\hat{\rm{B}}_{\rm {l}}^{\alpha} \, \hat{\rm{B}}_{\rm {l-1}}^{\alpha}
\cdots \hat{\rm{B}}_1^{\alpha} \,  \hat{\rm{B}}_{\Nt}^{\alpha} \cdots
\hat{\rm{B}}_{\rm {l+1}}^{\alpha}    \ .
\label{ADEF}
\end{eqnarray}

Summarizing, we have substituted the local fermionic interaction
in the partition function by an Ising model with a complicated 
multi-spin interaction. 
The remaining sum over the Ising field configurations in (\ref{partition}) 
can be computed by standard Monte Carlo techniques.
We adopt here the approach proposed by Blanckenbecler and coworkers 
\cite{Blancken1981},  extended to the Hubbard model at low temperatures 
in \cite{White1989,Gubernatis1991}. 

We refer to the above cited works for a discussion on the details of the
algorithm. In essence the update mechanism is based 
on the updating of the equal-time Green function
for an electron of spin $\alpha$ propagating through the 
field created by $\sigma(\rm{l})$
\begin{equation}
\hat{\rm{G}}^{\alpha}(\rm{l})_{\rm{ij}} \equiv \langle {\mathcal {T}} \,
[{\rm {c_{i\alpha} (\rm{l} \Dtau)}} \, {\rm {c_{j\alpha}^{\dagger} (\rm{l} \Dtau)}}] 
\rangle = [\id + \hat{\rm{A}}^{\alpha}(\rm{l}) ]_{\rm{ij}}^{-1} \ ,
\label{GDEF}
\end{equation}
with ${\mathcal {T}}$ denoting the temporal ordering operator.

The Green function turns out to be the fundamental object of the simulation
since it contains the information needed to update the field 
$\sigma (\rm{l})$. Besides, along the simulation,
observables like the energies and the local magnetic moment are calculated
as expectation values of certain matrix elements of $\hat{\rm{G}}({\rm {l}})$.

The computation of the Green function is unfortunately also the most expensive
part of the algorithm in terms of computing time.
The numerical evaluation of eq. (\ref{GDEF}) requires performing
$\Nt$ multiplications of matrices of dimension ${\rm {V}}$. That requires order
$\Nt \times {\rm {V}}^3$ operations plus the inversion of the resulting matrix,
which takes of order ${\rm {V}}^3$ operations.

Timings get worse at low temperature since
the matrices  ${\rm {\hat{\rm{B}}_{\rm{l}}}}$ 
get more and more ill-conditioned when increasing $\beta$. 
The computation of the product in eq. (\ref{GDEF}) is then plagued
with round-off errors. Obtaining a meaningful result
requires intermediate re-orthogonalizations in order to isolate
the divergent scales in the matrix product \cite{Gubernatis1991}.
For very large values of $\beta$ the Green function cannot even be 
calculated in a computer due to finite precision problems. 

The situation can be partially alleviated by realizing that eq. (\ref{GDEF})
immediately implies
\be
\hat{\rm{G^{\alpha}}}({\rm {l+1}}) = \hat{\rm{B^{\alpha}}}({\rm {l+1}}) \, 
\hat{\rm{G^{\alpha}}}({\rm {l}}) \, 
\hat{\rm{B^{\alpha}}}^{-1}({\rm {l+1}}) \ ,
\label{GADV}
\ee
which can be used to ``advance'' the Green function 
from time slice ${\rm {l}}$
to ${\rm {l+1}}$. The significant reduction in number of operations,
comes at the price of increasing the round-off errors. 
Due to this fact eq. (\ref{GADV}) can be used a limited number
of consecutive times, say till round-off errors become of 
the order of the statistical  ones. 
One then has to recompute $\hat{\rm{G}}({\rm {l}})$ according 
to eq. (\ref{GDEF}).
For the reasons discussed above, at low temperatures, say $\beta > 6$,  
the range of applicability of  eq. (\ref{GADV}) is very limited.

It is clear at this point that to accelerate the simulation
we have to concentrate efforts in speeding up matrix operations,
in particular the matrix multiplication. We address this point on more
quantitative grounds in the Appendix, where our implementation
of the algorithm in a cluster is also discussed.

%
%
%
%
 
\section{The Monte Carlo Simulation}

We have run numerical simulations on the general anisotropic Hubbard
model in ${\rm {d=3}}$
\bea
{\rm {\hat{H}}} =  & - \rm{t} &  \sum_{\langle \rm{i\,j} \rangle, \alpha}  \,
({\rm {c^{\dagger}_{\rm{i}\alpha}}} {\rm {c_{\rm{j}\alpha}}}  +
  {\rm {c^{\dagger}_{\rm{j}\alpha}}} {\rm {c_{\rm{i}\alpha}}}) 
 - \rm{tz} \, \sum_{\langle \rm{i\,j} \rangle, \alpha} \,
({\rm {c^{\dagger}_{\rm{i}\alpha}}} {\rm {c_{\rm{j}\alpha}}} +
  {\rm {c^{\dagger}_{\rm{j}\alpha}}} {\rm {c_{\rm{i}\alpha}}})  \nonumber \\
& +  {\rm {U}} & \, \sum_{\rm{i}}  ({\rm {n_{\rm{i}+}}} - 1/2)  \,
({\rm {n_{\rm{i}-}}} - 1/2)  \ .
\eea

In this notation $\rm{tz}$ and $\rm{t}$ represent
respectively the inter-planar and the in-plane hopping parameters. 
Varying $\rm{tz}$ the system undergoes a crossover between the purely
two dimensional behavior at $\rm{tz}=0$ and the three dimensional isotropic
case at ${\rm {t = tz}}$. Intermediate values of $\rm{tz}$  
model situations in which the material is better 
represented by a weakly coupled set of two dimensional layers, than by
a three dimensional isotropic lattice.

The phase diagram of the Hubbard model in $\rm{d=3}$ contains
a phase transition line in the plane $\beta \, - \, \rm{U}$. The high
temperature phase is paramagnetic while in the low temperature region
the ground state is an antiferromagnet with
the electron spins oriented staggered wise in all spatial directions.
The limiting behavior of the model for large \rm{U} corresponds to
the three dimensional quantum Heisenberg model. At $\rm{U=0}$ the system
becomes a gas of non-interacting electrons which shows no transition at all.

Along the numerical simulation we measure the kinetic 
and the Coulomb energies via the expectation value of 
the following operators
\bea
{\rm {e_k}} &=&\frac{{\mathcal {N}}}{4\rm{D}} \cdot 
\sum_{{\rm {i}}, \,\hat{\mu}} 
\langle {\rm {c^{\dagger}_{i \, \alpha}}}  
{\rm {c_{i+\hat{\mu} \, \alpha}}} \rangle \ , \\
{\rm {e_c}}& =&{\mathcal {N}} \cdot  
\sum_{{\rm {i}}} \langle {\rm {n_{i+}}}
{\rm {n_{i-}}} \rangle \ .
\eea
where the index $\hat{\mu}=0,\dots,5$ denotes the six spatial directions and 
the normalization factor ${\mathcal {N}} = 1 / ({\rm {V  \Nt}})$ accounting
for the sum over spatial and temporal lattices has been used.
We also measure the local magnetic moment defined as
\be
\rm{S^2} = {\mathcal {N}} \, \frac{3}{4} \langle 
( {\rm {n_{i+}}} - {\rm {n_{i-}}} )^2 \rangle \ .
\ee
Both, energies and magnetic moment, can be expressed as appropriate combinations
of matrix elements of the Green function.

\vspace{0.5cm}

The Ising variables are coupled to the \rm{z}-component of the electron
spin at each site. Taking this into account we construct the order parameter
in the following way. At each time slice \rm{l} we define
\be
{\rm {m^{l}_{stag}}} = \frac{1}{\rm{V}} \sum_{\rm{i}} 
(-1)^{\rm{x+y+z}} \cdot \sigma_{\rm{x\,y\,z}} \ ,
\ee
where {\rm  {(x,y,z)}} are the coordinates of site \rm{i}.
The average over configurations is defined by 
\be
{\rm {M^{l}_{stag}}} = \langle \,
\sqrt{ ( {\rm {m^{l}_{stag}}} )^2 } \, \rangle \ .
\ee
Adding up the contributions of all time slices we get
\be
\rm{M}_{{\rm {stag}}} = \frac{1}{\Nt} \sum_{\rm{l}} 
\rm{M}^{{\rm{l}}}_{{\rm {stag}}} \ .
\label{ORDER}
\ee

\begin{figure}[b!]
\begin{center}
\mbox{
\epsfig{figure=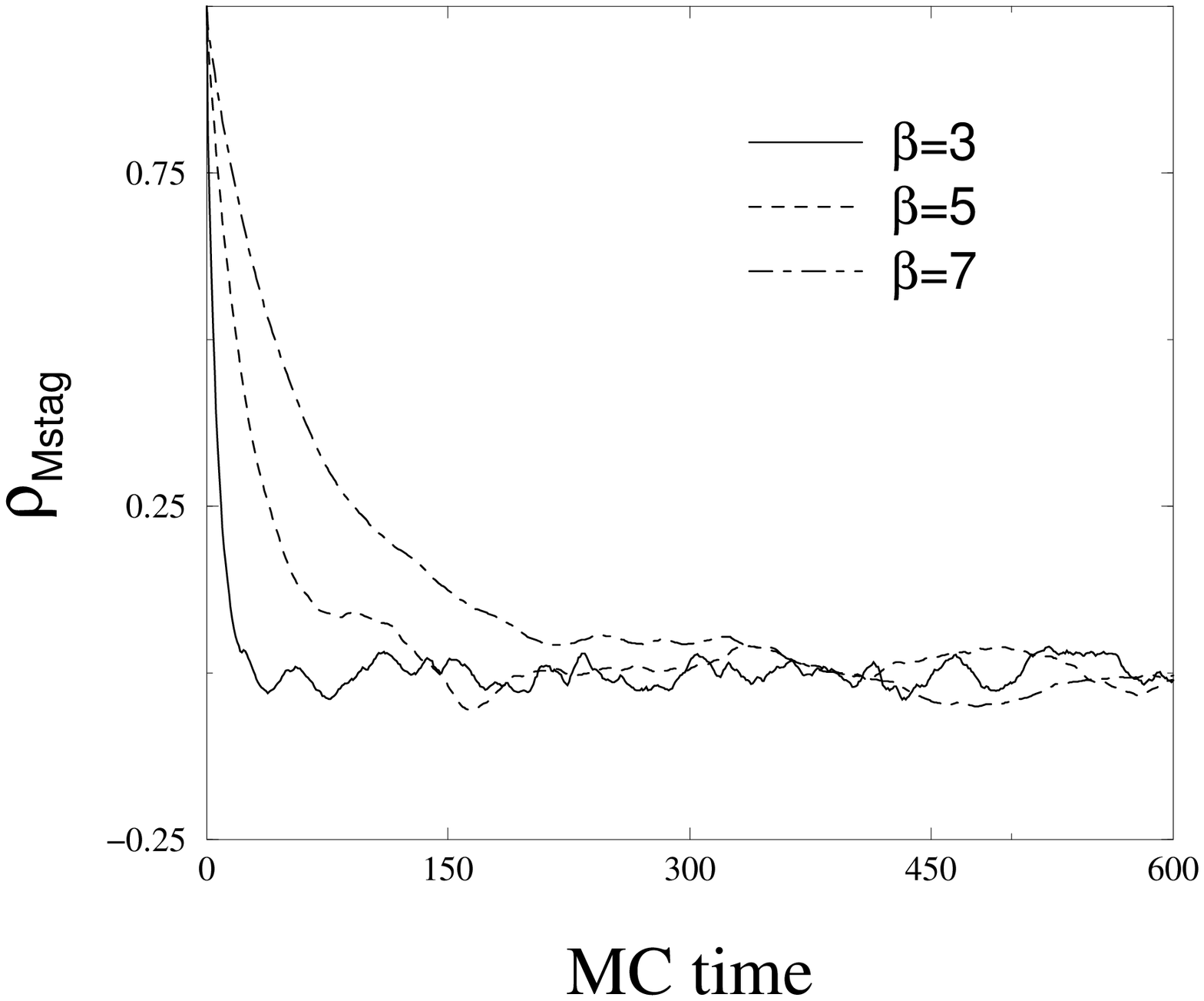,width=6cm,height=6.0cm}
\hspace{1em}
\epsfig{figure=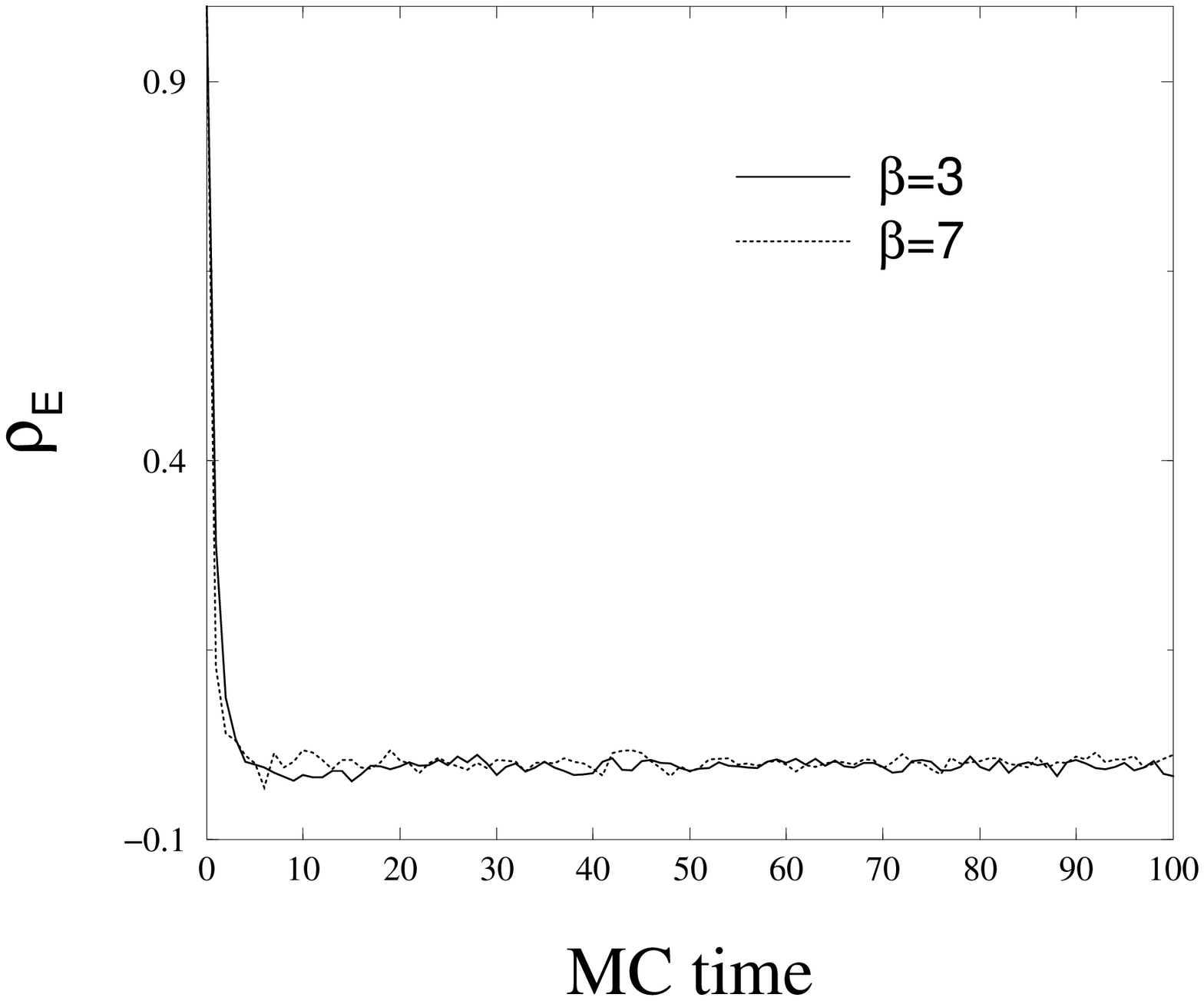,width=6cm,height=6.0cm}}
\end{center}
\vspace{-2.3em}
\caption[]{Autocorrelation function of the Staggered Magnetization
(a) and of the kinetic Energy (b). Note the change in the scale between
the two plots.}
\label{fig:l4corr}
\end{figure}

Along the simulation we have set $\rm{t} =1$. In this
paper we concentrate on the results for the isotropic case $\rm{t=tz}$.
Some exploratory studies at $\rm{tz < t}$ have also been performed.
In all cases
we work at fixed \rm{U} and sweep through $\beta$ looking for the value
at which the magnetic susceptibility has a maximum. 
To keep systematic errors under control we use 
always $\Dtau \sim 0.125$.
We are interested in the scaling properties of the susceptibility at the phase 
transition in order to measure the magnetic critical exponents.
For this purpose we concentrate most of our statistics at 
a single value of the Coulomb interaction, $\rm{U}=6$.
Here we run lattice sizes
${\rm{L}} =4,6,8,10$ with Monte Carlo times ranging from $10^5$ for
$\rm{L=4}$ to $10^4$ for $\rm{L=10}$ at each $\beta$ value. 
We discard between $10\%$ and $20\%$ of the data as thermalization time, 
depending on the parameter space point and the lattice size. 
The total computing time spent is the equivalent of about 240 months
in a Pentium III processor at \rm{1GHz}.

To assess the statistical quality of our data,
following \cite{SokalBOOK} we define the unnormalized autocorrelation 
function for the observable ${\mathcal {O}}$
\be
\rm{C}_{ {\mathcal {O}}}(\rm{t}) = 
\frac{1}{\rm{N-t}} \sum_{\rm{i=1}}^{\rm{N-t}}  
{\mathcal {O}}_{\rm{i}}  {\mathcal {O}}_{\rm{i+t}} - 
\langle  {\mathcal {O}} \rangle^2 \ .
\ee
as well as the normalized one
\be
\rho_{ {\mathcal {O}}}(\rm{t}) = \frac{\rm{C}_{ {\mathcal {O}}}(\rm{t})}
{\rm{C}_{ {\mathcal {O}}}(0)} \ .
\ee

The integrated autocorrelation time for  ${\mathcal {O}}$, 
$\tau_{ {\mathcal {O}}}^{\rm{int}}$, can be 
measured using the window method
\be
\tau_{ {\mathcal {O}}}^{\rm{int}}(\rm{t}) 
= \frac{1}{2} + \sum_{\rm{t^{\prime}} = 1}^{\rm{t}} 
\rho_{ {\mathcal {O}}}(\rm{t^{\prime}}) \ .
\ee
for large enough $t$, which is in practice selected self-consistently.
We use $\rm{t}$ in the range 5$\tau^{\rm{int}}$, 10$\tau^{\rm{int}}$, 
and we check that the obtained $\tau^{\rm{int}}$ remains stable as 
the window in $\rm{t}$ is increased.

In Figure \ref{fig:l4corr} we plot the autocorrelation function in $\rm{L=4}$
for the staggered magnetization (left side) 
and for the kinetic energy (right side). We observe the staggered magnetization
building up stronger autocorrelations as the phase transition is approached.
From this point of view the kinetic energy seems to be insensitive to 
changes in the temperature. We conclude from here that the order parameter
is a better observable to discuss the onset of criticality on the model.

%
%
%
%

\begin{figure}[htb!]
\begin{center}
\mbox{
\epsfig{figure=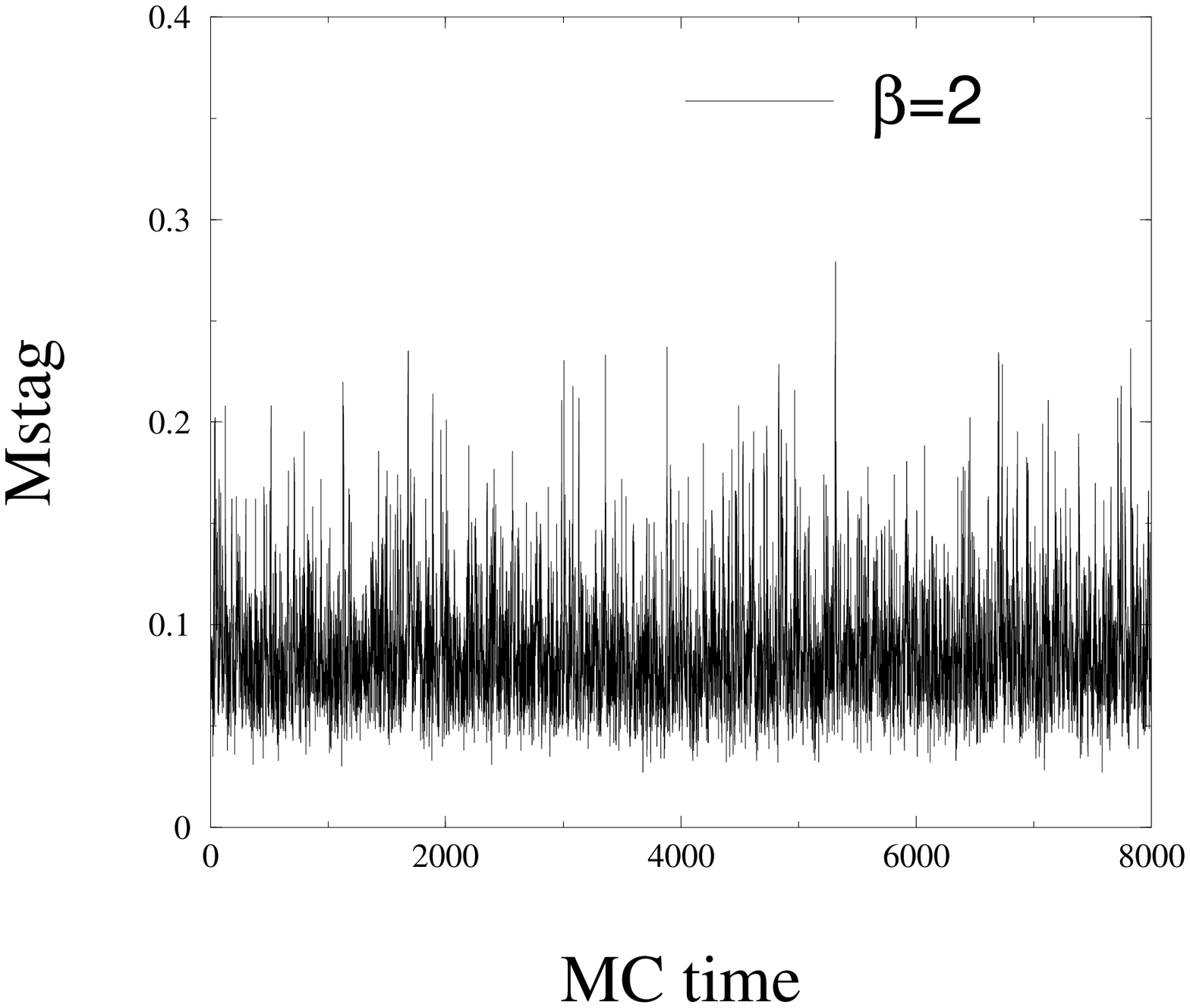,width=7cm,height=4.0cm}}
\end{center}
\begin{center}
\mbox{
\epsfig{figure=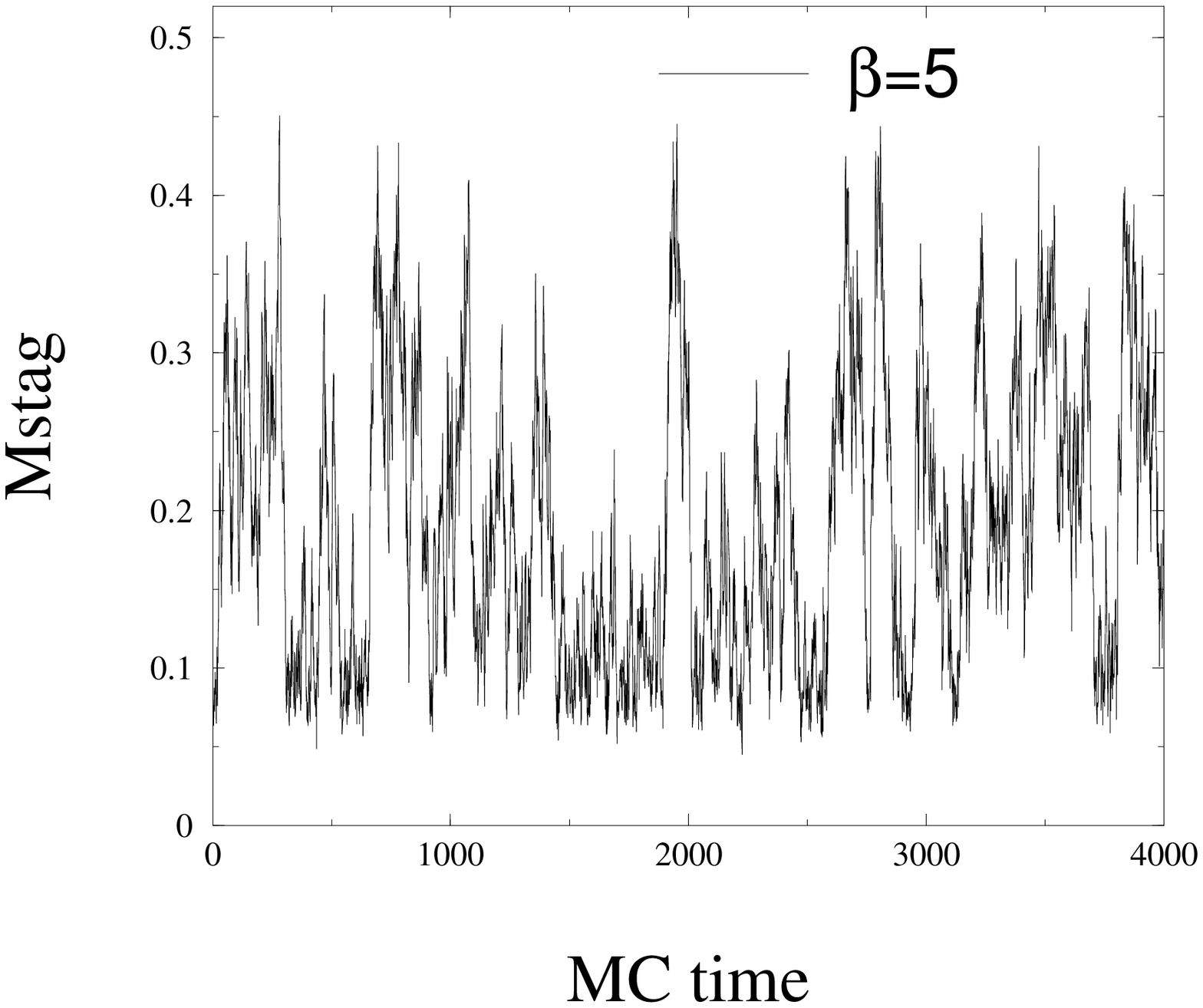,width=7cm,height=4.0cm}}
\end{center}
\vspace{-1em}
\caption[]{Monte Carlo evolution of ${\rm {M_{stag}}}$ at $\beta=2$
(upper part) and $\beta=5$ (lower part) for a $\rm{L=6}$ lattice 
at $\rm{U=6}$.}
\label{fig:evolution}
\end{figure}

\section{Isotropic Hubbard model}

We have run numerical simulations in lattices ranging from $\rm{L=4}$ 
to $\rm{L=10}$ to investigate the magnetic behavior of the system around the
N\'eel phase transition. For this purpose
we have measured the order parameter defined in eq. (\ref{ORDER})
and the staggered magnetic susceptibility
\be
\chi_{\rm {stag}} = \rm{V} \, \langle ({\rm {M_{stag}}})^2 \rangle \ ,
\label{SUSCEP}
\ee
which is a monotonically increasing function of $\beta$ for so is
$({\rm {M_{stag}}})^2$.

At low $\beta$ values, high temperatures, the system is in a disordered
paramagnetic phase. The mean value of 
the order parameter is zero up to corrections proportional to $\rm{1/V}$.
In figure \ref{fig:evolution}, upper part, we plot the MC evolution of 
${\rm {M_{stag}}}$ at $\beta=2$ in a $\rm{L=6}$ lattice with $\rm{U=6}$. 
As temperature decreases, and the magnetic phase transition
is approached, the value of the order parameter increases indicating
the tendency of the electron spins to organize themselves 
in a staggered way.
On the bottom of the same figure the MC evolution of 
${\rm {M_{stag}}}$ at $\beta=5$ in a $\rm{L=6}$ lattice at 
the same value of \rm{U} is plotted.
We see the system is flipping back and forth between the disordered 
paramagnetic phase and the staggered ordered one. There is a constant factor
$(1 - \rm{e}^{-\Dtau \rm{U}})^{-1}$ relating the two-point correlation functions
expressed in terms of the electron spin with the ones expressed in terms
of the Ising fields \cite{Hirsch1983}. Our plots of magnetic variables
contain already this factor.

\begin{figure}[htb!]
\begin{center}
\mbox{
\epsfig{figure=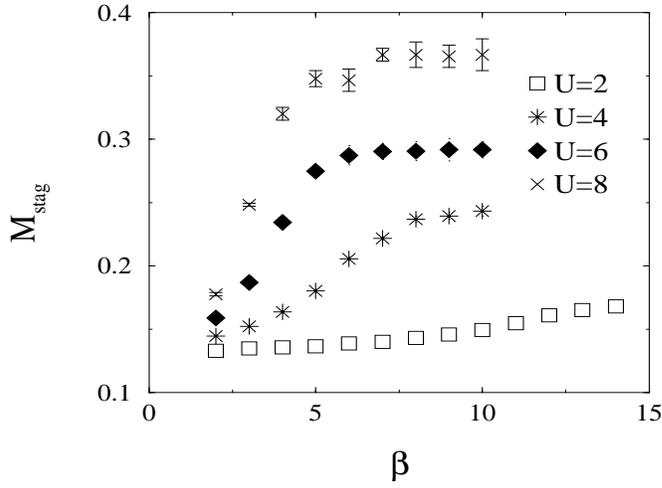,width=8cm,height=6.0cm}}
\end{center}
\vspace{-0.8em}
\caption[]{${\rm {M_{stag}}}$ versus $\beta$
for different values of $\rm{U}$ in $\rm{L=4}$.}
\label{fig:uvalues}
\end{figure}

\begin{figure}[htb!]
\begin{center}
\mbox{
\epsfig{figure=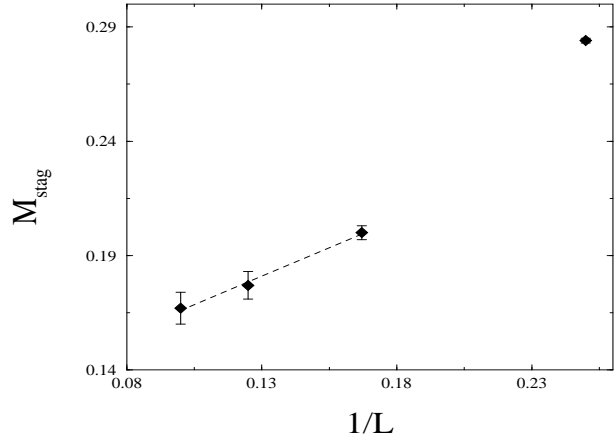,width=8cm,height=6.0cm}}
\end{center}
\vspace{-2.3em}
\caption[]{Asymptotic value of ${\rm {M_{stag}}}$ versus $\rm{1/L}$ 
at $\rm{U=6}$.}
\label{fig:mL}
\end{figure}

The staggered magnetization is an increasing function of $\beta$
for fixed $\rm{U}$. In fact, for all the lattice sizes we investigated
${\rm {M_{stag}}}$ increases with $\beta$ reaching a maximum
at some point, and developing a plateau afterwards. The onset of such 
plateaus is related to the maximal value that the magnetization 
can reach in that particular lattice size. 
In figure \ref{fig:uvalues} we display the results for an $\rm{L=4}$ lattice
for different values of the Coulomb interaction.
The values of the observables on the plateaus can be viewed as the 
asymptotic values in the $\rm{T=0}$ limit. 

In figure \ref{fig:mL} we plot those asymptotic values 
versus the inverse of the lattice size. Spin wave
theory predicts that the fluctuations giving rise to spin-spin correlations
decay as $\rm{1/L}$ \cite{Huse1988}.
Our results support this prediction for $\rm{L>4}$ giving
a value for ${\rm {M_{stag}}}$ in the thermodynamic limit of 0.156(3).
Including $\rm{L=4}$ the best fit is a $\rm{1/L^2}$ 
extrapolation which leads to a value of ${\rm {M_{stag}}}$ compatible with
the previous one. We can therefore not give a conclusive answer to this issue
but we are inclined to prefer the fit without the small 
lattice because finite size effects are likely to be uncontrolled for 
$\rm{L=4}$. Indeed, the asymptotic value of 
${\rm {(M_{\rm{stag}})^2}}$ starts to stabilize only from $\rm{L=6}$ on.

Next, we focus on the critical behavior of the system. We concentrate our
largest statistics in this particular aspect. Our aim is studying the
scaling of the order parameter and the susceptibility close to the
phase transition temperature, and finally extracting the magnetic critical
exponents. A first question arising is the actual value of 
the transition temperature.
The most recent work the authors are aware of \cite{Muramatsu2000} 
quotes a value $\rm{T_c \sim 0.3}$ for the N\'eel transition at \rm{U=6}.
This temperature is certainly much lower than the ones reported in 
the pioneer works of the $1980$s (see eg. \cite{Scalapino1989}). 
Our purpose is to give an estimation of the critical temperature
based on the measurements of the order parameter.

\begin{figure}[htb!]
\begin{center}
\mbox{
\epsfig{figure=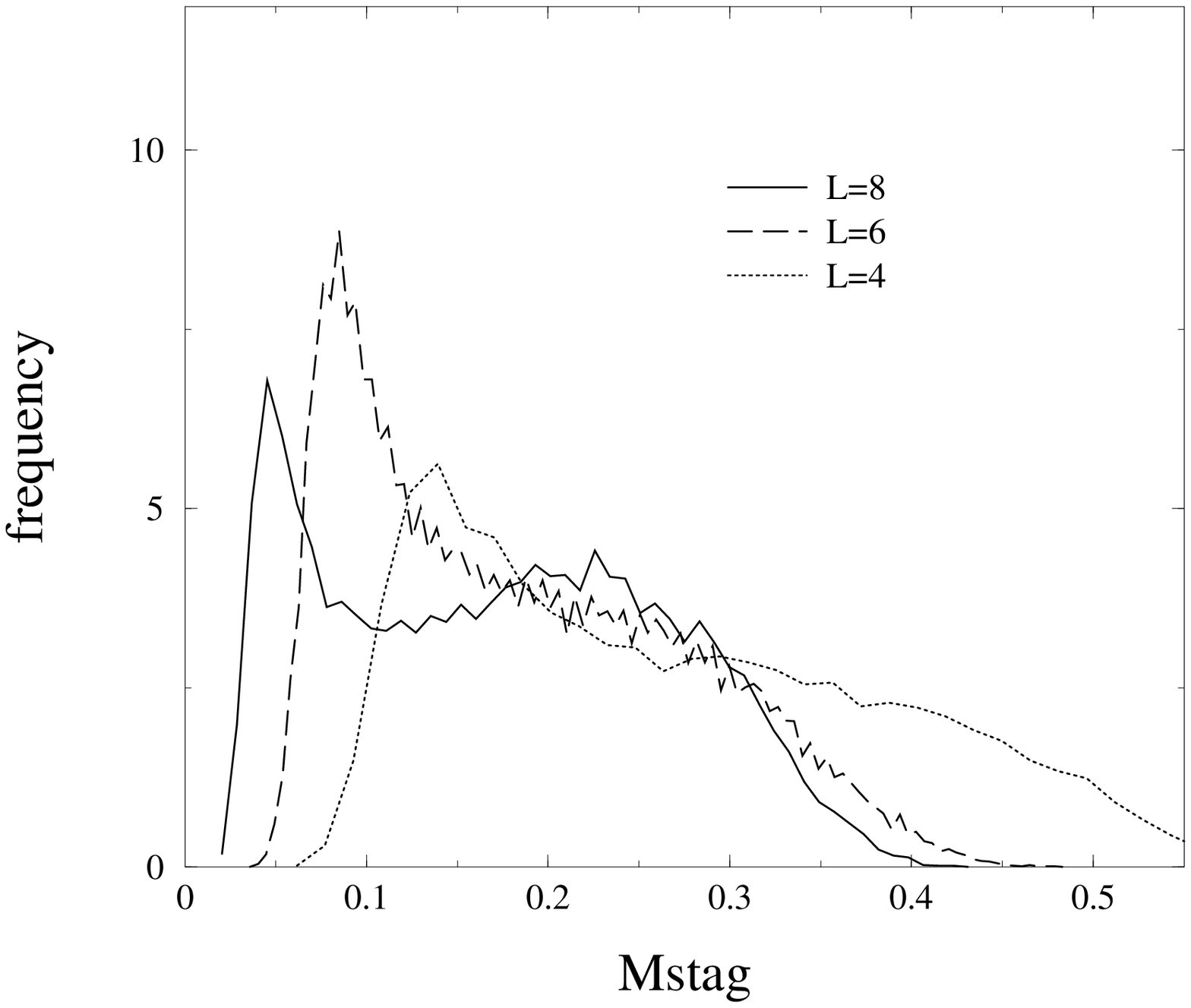,width=8cm,height=6.0cm}}
\end{center}
\begin{center}
\mbox{
\epsfig{figure=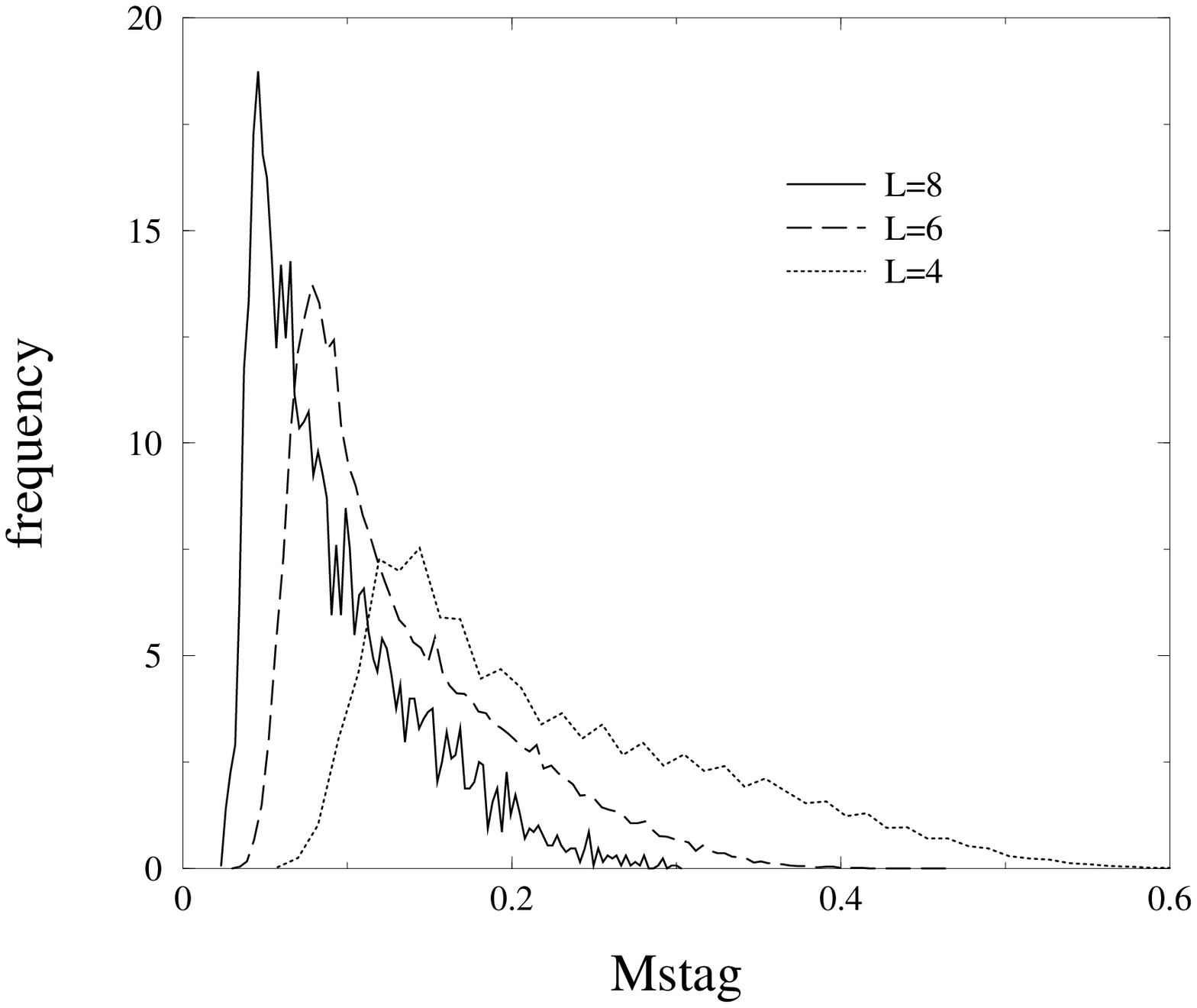,width=8cm,height=6.0cm}}
\end{center}
\vspace{-2.3em}
\caption[]{Normalized distribution of ${\rm {M_{stag}}}$ 
at $\rm{U=6}$ at $\beta=4$ (lower plot) and
$\beta=5$ (upper plot). The N\'eel phase transition takes
place between this two values.}
\label{fig:magnetization}
\end{figure}

\begin{figure}[htb!]
\begin{center}
\mbox{
\epsfig{figure=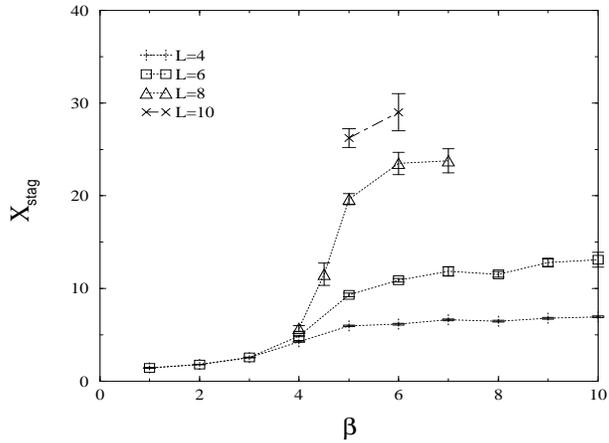,width=8cm,height=6.0cm}}
\end{center}
\vspace{-2.3em}
\caption[]{$\chi_{\rm{stag}}$ versus $\beta$ at $\rm{U=6}$
and lattice sizes $\rm{L=4,6,8}$ and $10$.}
\label{fig:susceptibility}
\end{figure}

In figure \ref{fig:magnetization} we plot the histograms corresponding
to the time evolution of ${\rm {M_{stag}}}$ for increasing lattice
sizes at $\beta=4$ (lower plot) and $\beta=5$ (upper plot).  
The asymmetry of the distributions at $\beta=4$ indicates that we are close
to a phase transition. However the system is still clearly
in the paramagnetic side because when increasing the lattice size the 
peak of the magnetization in the paramagnetic region tends to dominate 
the distribution while the other runs away.
The behavior is radically different at $\beta=5$. The peak corresponding 
to the paramagnetic phase decreases when the lattice size gets bigger,
corresponding therefore to a finite size effect. Conversely, the peak
in the symmetry broken phase tends to grow. Summarizing,
the system is already in the antiferromagnetic N\'eel phase 
at $\beta=5$ because for increasing lattice size the system stabilizes 
in the staggered phase.

From the order parameter distributions it is clear that the magnetic 
transition takes place between $\rm{T=0.25}$ and $\rm{T=0.20}$. Our results
therefore support the ones of Muramatsu and coworkers in the sense
that the phase transition occurs at a value much lower than traditionally
expected.  The fact that we measure a value even slightly smaller
might be related to the use of a different observable. In \cite{Muramatsu2000}
the authors use cumulants of the energy to locate the critical point,
while our results are based on order parameter measurements. 
In principle, different estimators give slightly 
different results in finite lattices.

\begin{figure}[htb!]
\begin{center}
\mbox{
\epsfig{figure=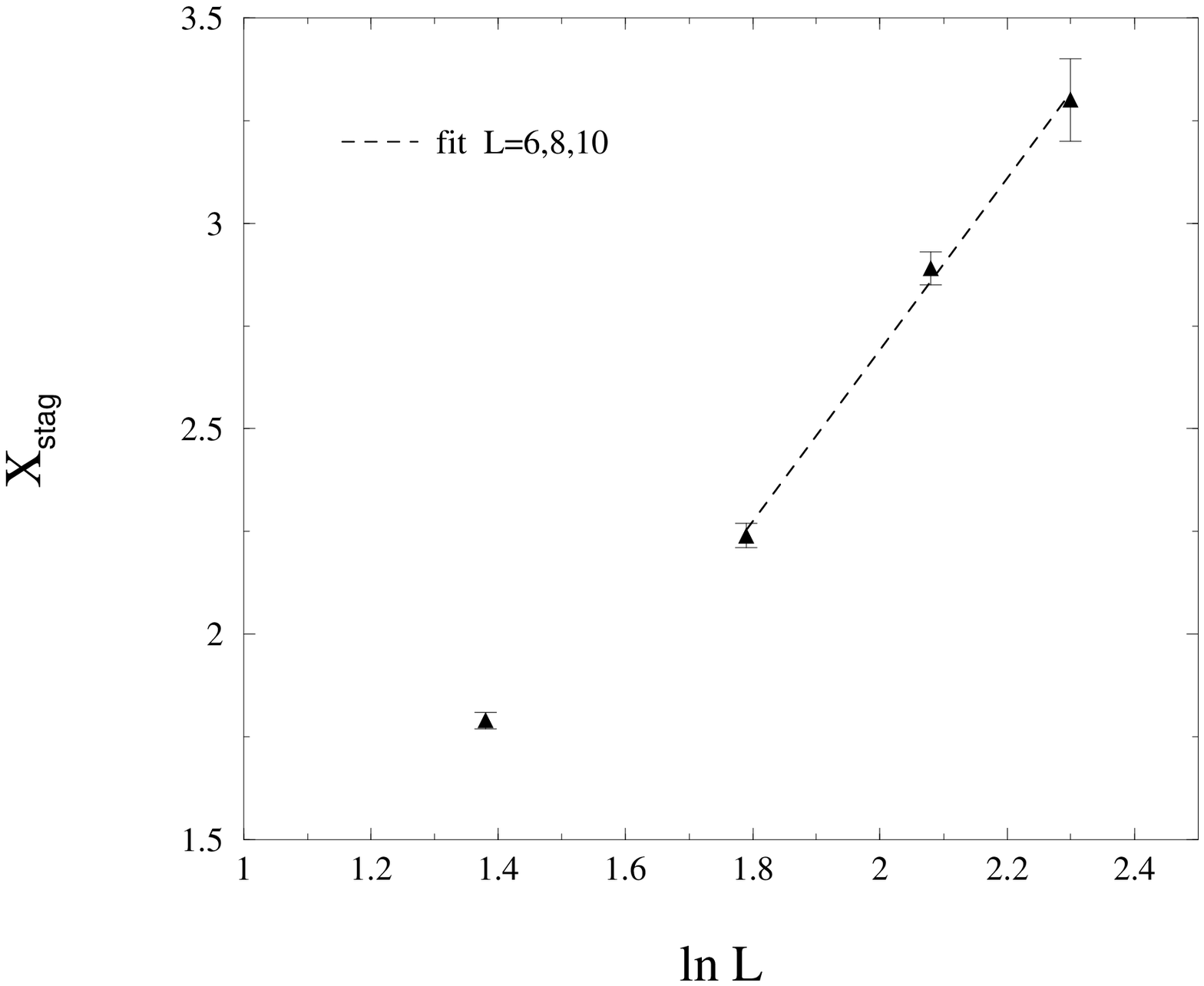,width=8cm,height=6.0cm}}
\end{center}
\vspace{-2.3em}
\caption[] {\rm{Log-log} plot of the magnetic susceptibility
at $\beta=5$, \rm{U=6} versus lattice size. The linear fit gives a
critical exponent $\gamma/\nu = 2.08(9)$.}
\label{fig:gamma}
\end{figure}

The behavior of the staggered susceptibility (eq. \ref{SUSCEP}) 
across the parameter space for different lattice sizes is presented
in figure \ref{fig:susceptibility}. We observe the susceptibility growing
monotonically until reaching a plateau reflecting the behavior of the 
magnetization itself. The saturation of the magnetization,
and the subsequent plateau in the susceptibility raises doubts on
the interpretation of the peak in the susceptibility
as a good observable to locate the phase transition.
In principle, there is no reason to believe that the $\beta$ value at which
the susceptibility reaches the plateau has anything to do with the critical
temperature of the N\'eel transition.

We observe that the onset of the plateaus comes close to but 
clearly after $\beta=5$
in all cases. Taking $\beta=5$ as our best estimate for the
critical temperature, and using the scaling law
\be
\chi_{\rm {stag}}(\rm{T_c}) \propto \rm{L^{\gamma/\nu}} \ ,
\label{gamma}
\ee
our result for the magnetic critical exponent should
agree with the one of the three dimensional Heisenberg model 
\cite{Sandvik1998}, that is, $\gamma/\nu \approx 1.98$.
Our estimation for the magnetic critical exponent is in good agreement
with this expectation. The result of our linear fit is plotted
in figure \ref{fig:gamma}, giving a value of $\gamma/\nu = 2.08(9)$.

%
%
%
%

\section{Anisotropic Hubbard model}

The introduction of an anisotropic hopping parameter \rm{tz}
allows us to interpolate between the purely two dimensional behavior
($\rm{tz=0}$) and the perfectly isotropic three dimensional lattice
($\rm{tz=t}$). We have done some exploratory studies at
intermediate values of  \rm{tz} to get some insight on the
crossover behavior of the model.

\begin{figure}[htb!]
\begin{center}
\mbox{
\epsfig{figure=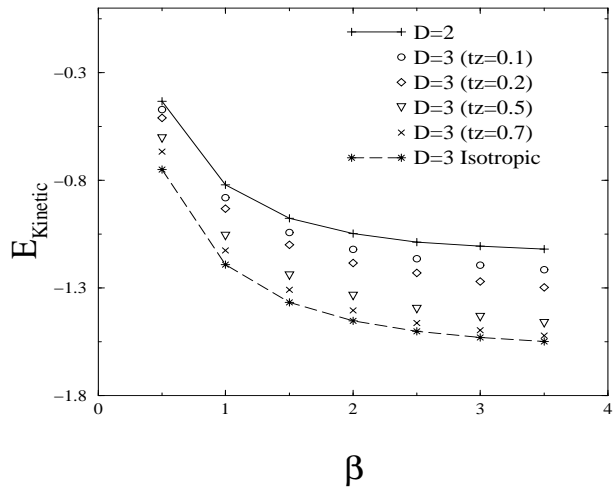,width=8cm,height=6.0cm}}
\end{center}
\vspace{-1em}
\caption[]{Kinetic energy for different values of \rm{tz} at \rm{U=4}. 
The values of the energies interpolate smoothly between 
the \rm{D=3} and \rm{D=2} limits.}
\label{fig:ener_ani}
\end{figure}

In figure \ref{fig:ener_ani} we plot the result for the kinetic
energy in $\rm{D=2,3}$ and at several intermediate values of the
hopping parameter. The interpolation is smooth in all cases. 
As a function of \rm{tz}, the interpolation is linear for the high temperature
case, and faster than linear when the temperature is lowered.
This can be understood easily taking into account that at low temperatures
the correlation between planes is higher and compensates the
smaller value of the interplanar \rm{tz} hopping coupling.

In two dimensions a phase transition to an antiferromagnetic
ordered phase is expected at $\rm{T=0}$. In principle we can ask
for the dependence of the phase transition temperature on the
anisotropic hopping parameter \rm{tz}. Evidently decreasing
\rm{tz} will result in a decrease of the transition temperature.
The actual dependence $\rm{T_c (tz)}$ is an interesting 
monitor of the form of the quantum fluctuations which disorder
the ground state at $\rm{T=0}$ in $\rm{D=2}$. 

\begin{figure}[ht!]
\begin{center}
\epsfig{figure=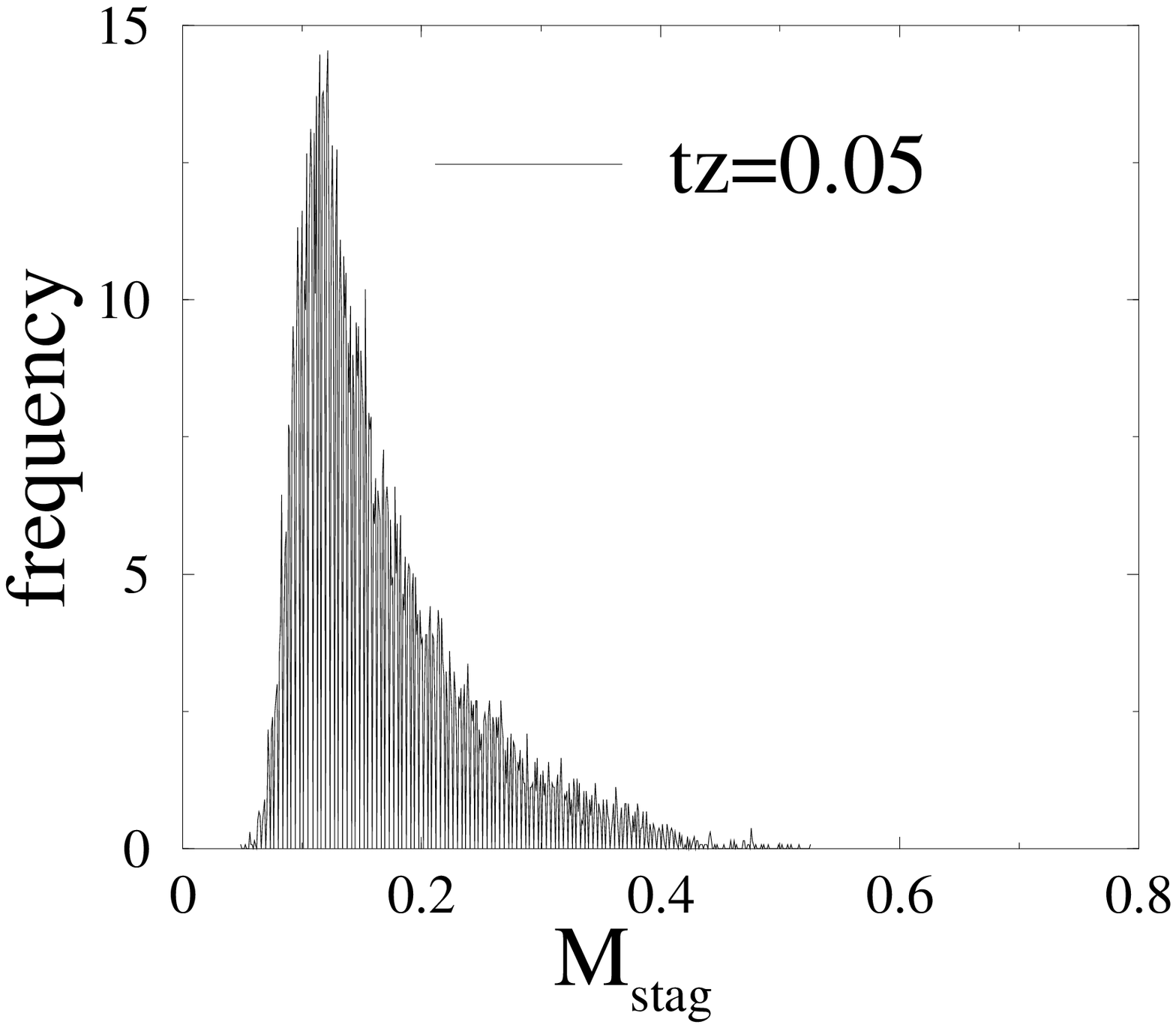,width=5cm,height=4.0cm}
\hspace{0.5cm}
\epsfig{figure=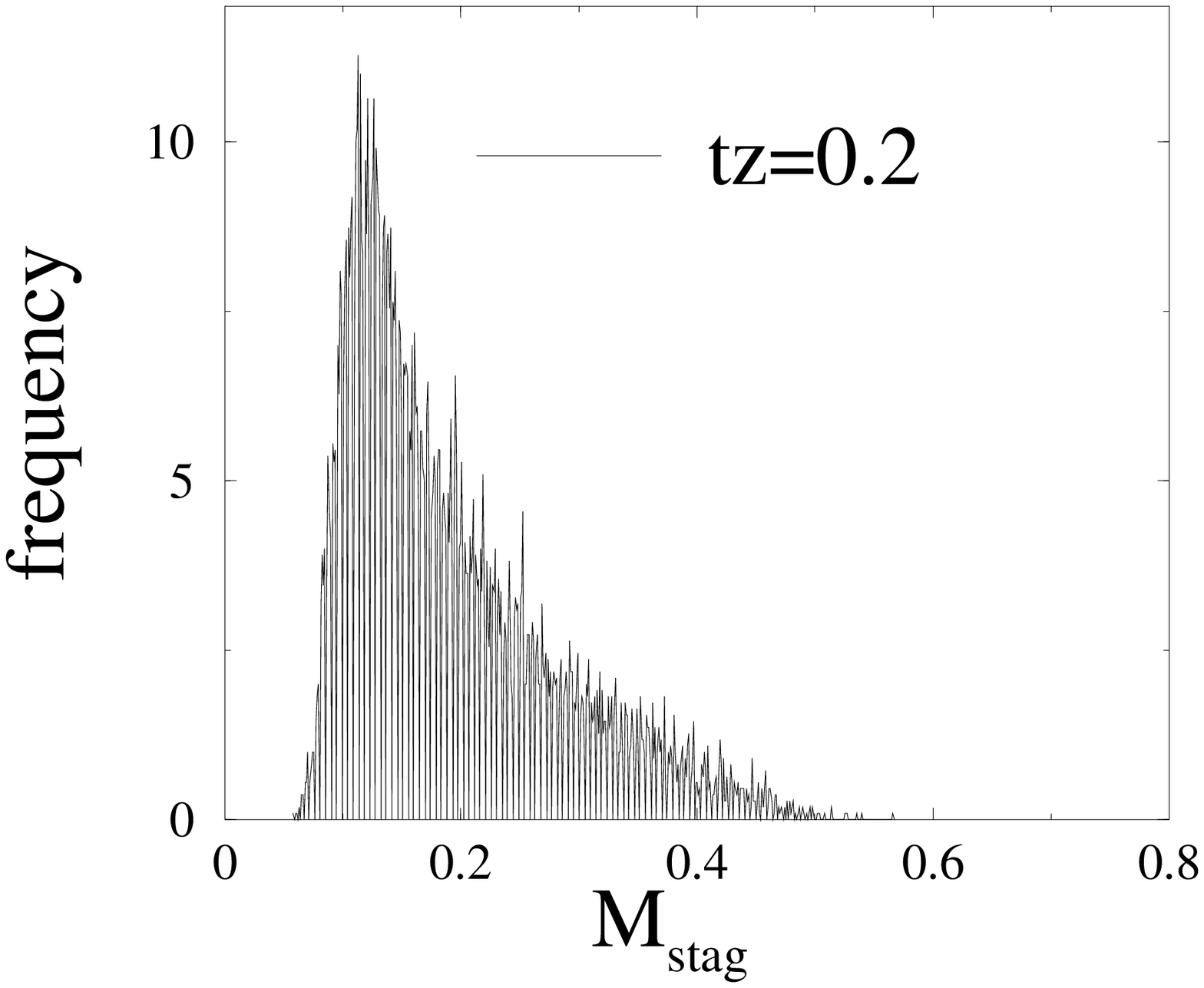,width=5cm,height=4.0cm}
\vspace{0.5cm}
\epsfig{figure=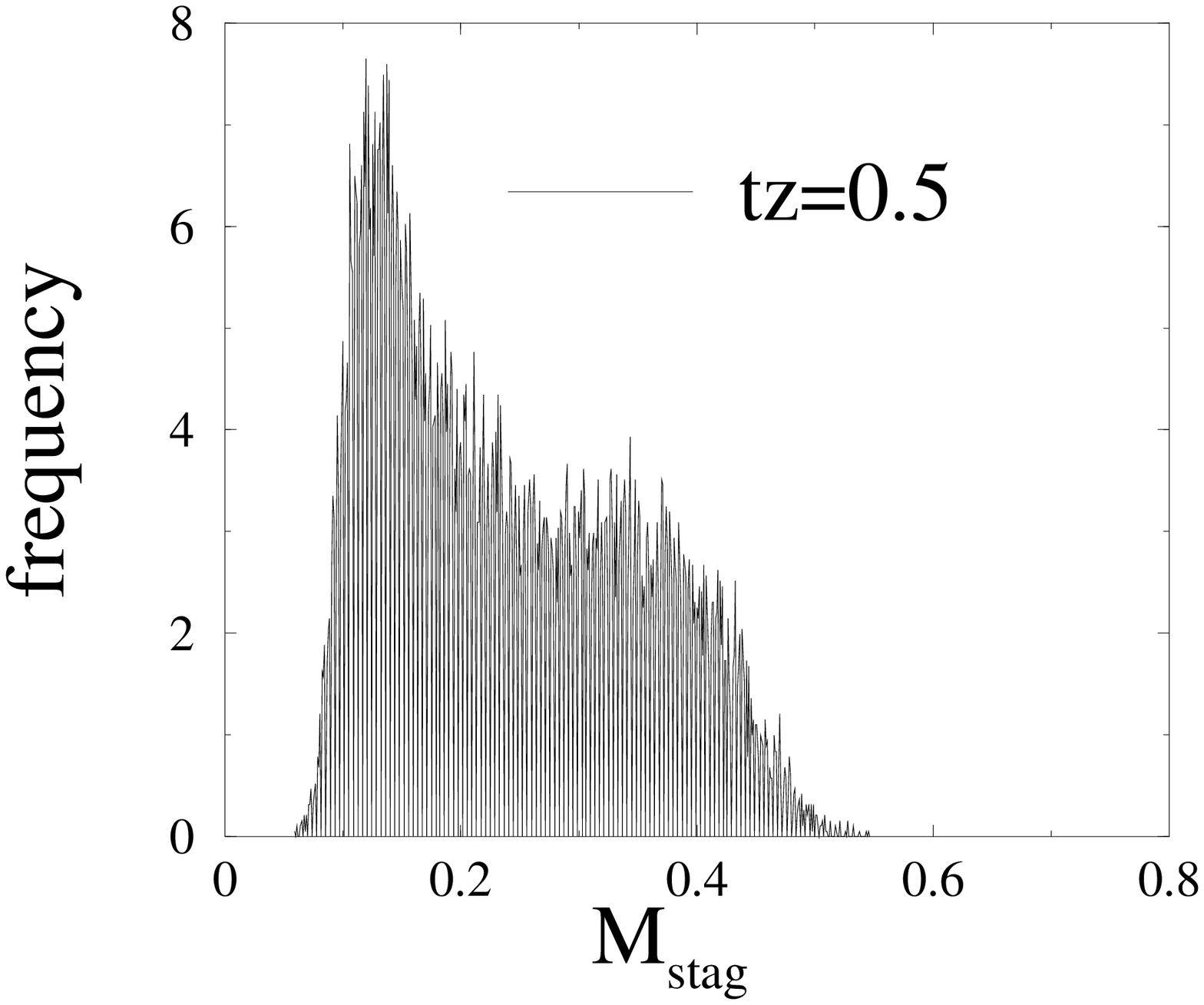,width=5cm,height=4.0cm}
\hspace{1cm}
\epsfig{figure=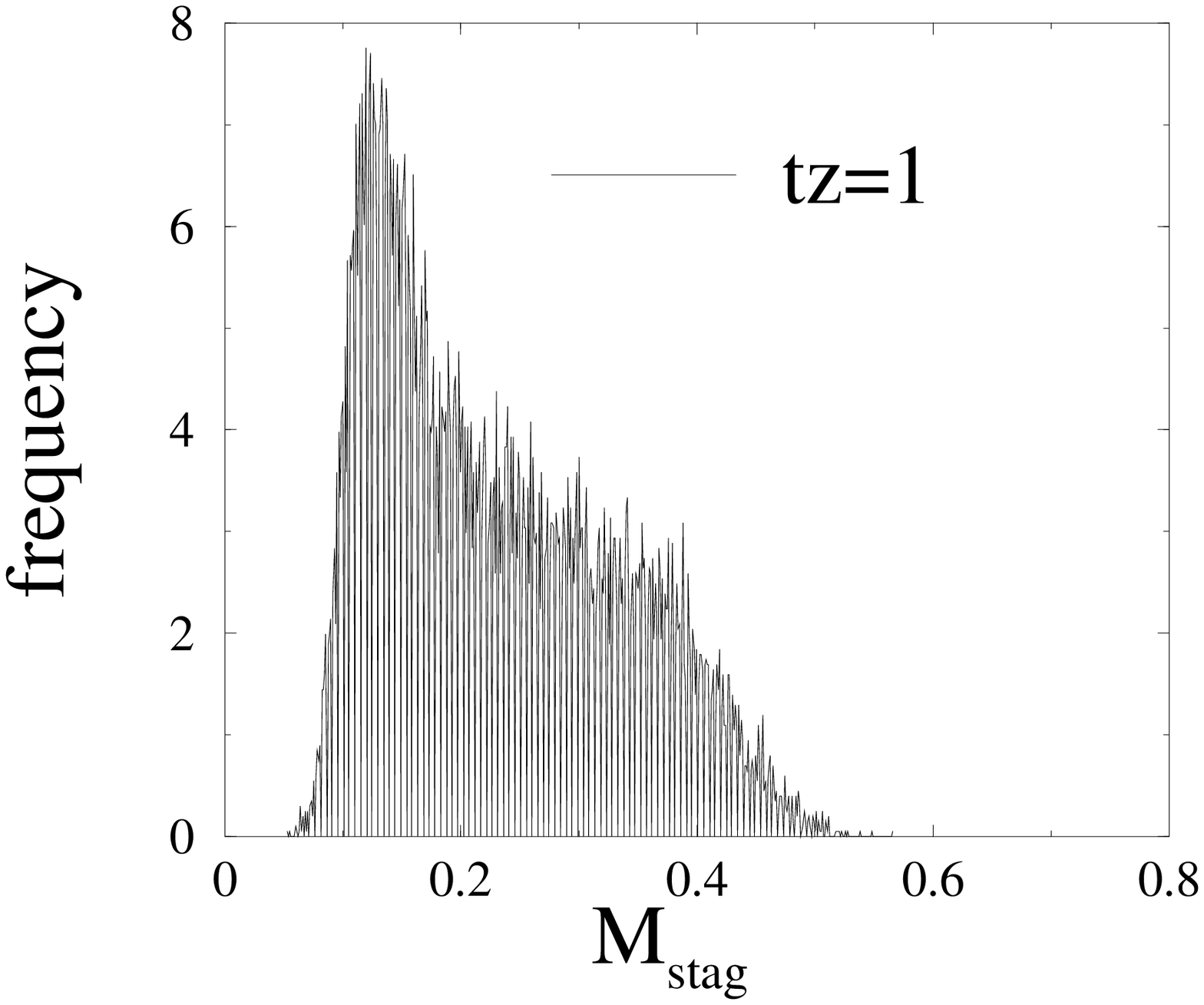,width=5cm,height=4.0cm}
\end{center}
\vspace{-2.3em}
\caption[]{Distribution of ${\rm {M_{stag}}}$ at $\beta=6$
in a $\rm{L=4}$ lattice at $\rm{U=6}$ for several values of \rm{tz}.}
\vspace{-1em}
\label{fig:combinedtzs}
\end{figure}

As a first step in that direction we have measured the staggered
magnetization for a small lattice, $\rm{L=4}$, for different
values of \rm{tz}. 
The order parameter flips between the disordered phase and the ordered
one at $\beta=6$ producing these distributions plotted 
in figure \ref{fig:combinedtzs}. 
We can observe that as \rm{tz} is lowered, the system is the more 
and more in the disordered phase, meaning that the transition 
temperature in fact decreases
as \rm{tz} goes to zero. Eventually, the value of the critical temperature
should go to zero, where the quantum critical point of the 
$\rm{D=2}$ Hubbard model is expected to be.

%
%
%
%
 
\section{Summary and conclusions}

We have investigated the properties of the magnetic phase
transition in the three dimensional Hubbard model.
The measurement of the order parameter allows us to give an
estimation of the critical temperature at $\rm{U=6}$. From
the scaling of the magnetic susceptibility we compute the magnetic
critical exponent $\gamma/\nu$ which is in agreement with the
magnetic exponent of the three dimensional Heisenberg model.

The dependence of the phase transition temperature on the
anisotropic hopping parameter is a very interesting
project from the numerical point of view. In this area
we are aware of results based on Dynamical Mean Field 
Theory and the Two Particle Self Consistent approach\cite{Marie1996}. 
It would certainly be of interest to probe
such results in a Monte Carlo simulation.

We have shown that computationally costly thermodynamic quantities 
such as distributions of the order parameter and
critical exponents can be computed with moderate-large computing 
resources using a well known algorithm. This fact should not
dismiss a very important issue, which is the development of better
core algorithms. 
A consequence of the impressive development of computer technologies
is that we are now able to produce results that we could not have dreamed
of only 5 years ago. Such technical developments should go hand by hand
with work in algorithm improvement.

\vspace{1cm}
{\bf {Acknowledgments:}} I.C. is grateful to J. Gubernatis for his assessment
in questions regarding low temperature simulations
and to V. Mart\'{\i}n-Mayor for enlightening discussions.
J.W.D. wishes to acknowledge discussions with
J.~Gubernatis and J. Glimm. Financial support has been provided by
the USA Department of Energy, under contract number DE-AC02-98CH10886.

\newpage

\section*{Appendix: The Hubbard model on a PC cluster}

The question we try to answer here is how to make
optimal use of the PC cluster to speed up
our numerical investigation of the Hubbard model using
the determinantal method.
As we pointed out in the introduction the farm method might not
been suitable to simulate large lattices due to thermalization issues.
The answer to our question is probably that
a combination of both, the {\sl {farm method}}, and a parallel 
version of the algorithm would do best.

The most serious problem regarding parallelization techniques
applied to this algorithm is the extreme non-locality of the update
mechanism.
Let us suppose that we distribute the Ising spin variables
among {\sl {np}} processors in such a way that each processor
takes care of the update of a piece of the spatial lattice.
It is easy to show that such strategy could not work. 
Consider for instance the update of the spin $\sigma_{\rm{s}}$ 
(we will omit the time index {\rm {l}} throughout this section for
notation clarity)
The update probability depends, among other things,  
on the diagonal element of the Green function $\rm{G}^{\alpha}_{\rm {ss}}$.
The crucial point is that if the spin $\sigma_{\rm {s}}$ is flipped
all the elements of the Green function change \cite{White1989}
\bea
& \rm{G}^{\alpha}_{\rm{ij}} & \rightarrow  {\rm {f}} \, 
\rm{G}^{\alpha}_{\rm{ij}} + \rm{G}^{\alpha}_{\rm{is}} \cdot \rm{G}^{\alpha}_{\rm{sj}} \qquad
\qquad \quad \qquad \mbox{\rm{if \quad j $\neq$ s}}  \nonumber \\ 
& \rm{G}^{\alpha}_{\rm{is}} & \rightarrow \rm{G}^{\alpha}_{\rm{is}} + 
{\rm {f}} \, (\rm{G}^{\alpha}_{\rm{is}} \cdot \rm{G}^{\alpha}_{\rm{ss}} 
- \rm{G}^{\alpha}_{\rm{is}}) \quad \qquad \mbox{\rm{if \quad j=s}}  \ ,
\label{UPDATE}
\eea
where ${\rm {f}}$ is a constant factor independent of the spatial coordinates.
In particular the change affects all diagonal elements which
enter in the update probability of all the other $\sigma$'s on the 
the processors. This implies that every time a spin is flipped
the recomputed Green function should be communicated to all processors.
Synchronizing such communication is likely to be impossible. In any case,
from a strictly performance point of view it is clear such an strategy
could not pay off. The update of the Ising fields
within a same time slice must therefore be sequential. 

A look at the definition of the Green function (eq. (\ref{GDEF})) tells us
that the update of the different time slices cannot be a distributed
task either. The value of the Green function at a particular
time slice {\rm {l}}, depends on the state of the Ising spins on all
the other time slices. 
From the previous analysis we conclude that the update
process is inherently sequential in all dimensions.

A possibility to still make use of computing cooperation among several
processors is to parallelize the matrix operations \cite{Para2002}.
We observe in eq. (\ref{UPDATE})
that despite all the matrix elements $\rm{G}^{\alpha}_{\rm {ij}}$ of the
Green function change after a spin flip, such change can be
computed from the original $\rm{G}^{\alpha}_{\rm {ij}}$ plus a 
factor which only depends on the elements of $\rm{G}^{\alpha}$ belonging 
to the row and column of the particular site $\rm{s}$ being updated
\begin{center}
$
\rm{\hat{G}}^{\alpha} = \left( 
\begin{array}{ccccc}
\cdots & \cdots & {\rm{G}^{\alpha}_{\rm{1s}}} & \cdots & \cdots   \\
\cdots & \cdots & \cdots & \cdots & \cdots \\
\cdots & \cdots & {\rm{G}^{\alpha}_{\rm{s-1}\rm{s}}} & \cdots & \cdots \\
{\rm{G}^{\alpha}_{\rm{s1}}} & \cdots & {\rm{G}^{\alpha}_{\rm{ss}}} & \cdots & 
\rm{G}^{\alpha}_{\rm{sV}} \\
\cdots & \cdots & {\rm{G}^{\alpha}_{\rm{s+1s}}} & \cdots & \cdots \\
\cdots & \cdots & \cdots & \cdots & \cdots \\
\cdots & \cdots & {\rm{G}^{\alpha}_{\rm{Vs}}} & \cdots & \cdots \\
\end{array}
\right)
$
\end{center}

The parallelization strategy takes profit of this regularity.
The matrix elements of the different operators are distributed column-wise
among the processors. When a matrix operation takes place, eg. a matrix
multiplication, each processor computes only the part corresponding to
the column it is responsible for. 
The update of the field $\sigma_{\rm{s}}$ is done simultaneously
on all the processors. In principle, since the random number sequence is the
same for all of them, the result of the update is the same on all 
of them. Therefore this mechanism, although redundant, is harmless.
In the program, before the update function is called,
the processor containing the column $\rm{G}^{\alpha}_{\rm{is}}$
broadcast this column to all the others.
As explained before this is the only information needed to recompute
the Green function during the update process.
Note that the portion of the \rm{s}-row needed by each processor
is stored locally, and therefore needs not to be communicated.

\begin{figure}[htb!]
\begin{center}
\mbox{
\epsfig{figure=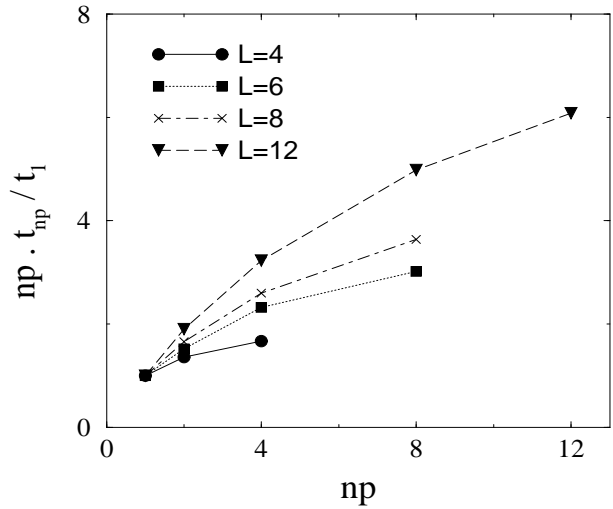,width=8cm,height=7.0cm}}
\end{center}
\vspace{-1em}
\caption[]{Speed up of the simulation as a function of the number
of processors using the parallelization method described in the text.}
\label{fig:speed}
\end{figure}

In figure \ref{fig:speed} we show the speed up in the calculation of
the Green function when using the proposed strategy. 
For small and intermediate lattice sizes the {\sl {farm method}}
is still the best option. For big lattices ($\rm{L\geq10}$)
the algorithm starts to scale reasonably well. 
It is clear that there is a bottleneck generated by the communication
of matrix elements during the calculation. The jam tends to improve when
the lattice size gets big because the processors have more operations
to perform and therefore do not block the channels trying to submit
and retrieve information constantly. 

Summarizing, the lesson to extract from here is that the algorithm 
is parallelizable. Using this scheme instead of a
{\sl {farm method}} pays off for big lattices. 
It is also clear that we have a very modest switch (Fast Ethernet at 
100 Mbits/s).The performance of the parallelization is probably 
boosted using a more advanced switch for instance a Mirinet$^{\rm{TM}}$
\cite{MIRINET}.


\newpage



\begin{thebibliography}{20}
 


\bibitem{Hubbard1963}
J. Hubbard
{\sl Proc. Roy. Soc. London} {\bf A276} p 238 (1963)


\bibitem{White1989}
S.R. White et al.
{\sl Phys. Rev.} {\bf B40} p 506 (1989) 


\bibitem{Gubernatis1991}
E.Y. Loh and J.E. Gubernatis in
{\sl ``Electronic Phase Transitions'', Ed. W. Hanke and Yu V. Kopaev} p 177
(1992)

\bibitem{Hirsch1987}
J.E. Hirsch
{\sl Phys. Rev.} {\bf B35} p 1851 (1987)

\bibitem{Scalapino1989}
R.T. Scalettar, D.J. Scalapino, R.L. Sugar and D. Toussaint
{\sl Phys. Rev.} {\bf B39} p 4711 (1989)


\bibitem{Muramatsu2000}
R. Staudt, M. Dzierzawa, A. Muramatsu
{\sl Eur. Phys. J.} {\bf B17} p 411 (2000)


\bibitem{Sandvik1998}
A.W. Sandvik
{\sl Phys. Rev. Lett.} {\bf 80} p 5196 (1998)



\bibitem{Martin2001}
Martin L\"uscher
{\sl ``Lattice QCD on PCs?''} {\bf hep-lat/0110007} \\
{\sl Proceedings Lattice 2001,}
Elsevier (to appear)
 

\bibitem{Blancken1981}
R. Blanckenbecler et al.
{\sl Phys. Rev.} {\bf D24} p 2278 (1981)
 



\bibitem{Hirsch1983}
J.E. Hirsch,
{\sl Phys. Rev.} {\bf B28} p 4059 (1983)

\bibitem{SokalBOOK}
A.D. Sokal 
{\sl Bosonic Algorithms} in {\sl Quantum Fields on the Computer}.
Advanced Series on Direction in High Energy Physics Vol. 11, M. Creutz Editor.
World Scientific, Singapore 1992.

\bibitem{Huse1988}
D.A. Huse
{\sl Phys. Rev. B} {\bf 37} p 2380 (1988)


\bibitem{Marie1996}
A.M. Dar\'e, Y.M. Vilk and A.-M.S. Tremblay
{\sl Phys. Rev. B} {\bf 53} p 14236 (1996)





\bibitem{Para2002}
I. Campos, J.W. Davenport and Wonho Oh
(Unpublished)

\bibitem{MIRINET}
{\bf http://www.mirinet.com} 

\end{thebibliography}
\end{document}